\documentclass[conference]{IEEEtran}
\IEEEoverridecommandlockouts
\usepackage{cite}
\usepackage{amsmath,amssymb,amsfonts}
\usepackage{algorithmic}
\usepackage{graphicx}
\usepackage{textcomp}
\usepackage{xcolor}
\usepackage{url}
\usepackage{listings}
\def\BibTeX{{\rm B\kern-.05em{\sc i\kern-.025em b}\kern-.08em
    T\kern-.1667em\lower.7ex\hbox{E}\kern-.125emX}}
\begin{document}

\title{Is Rust C++-fast? Benchmarking System Languages on Everyday Routines}

\author{\IEEEauthorblockN{Nikolay Ivanov}
\IEEEauthorblockA{\textit{Dept. of Computer Science and Engineering} \\
\textit{Michigan State University}\\
East Lansing, MI, USA \\
ivanovn1@msu.edu}
}

\maketitle

\begin{abstract}
Rust is a relatively new system programming language that has been experiencing a rapid adoption in the past 10 years. Rust incorporates a memory ownership model enforced at a compile time. Since this model involves zero runtime overhead, programs written in Rust are not only memory-safe but also fast, leading to performance comparable to C and C++. Multiple existing benchmarks comparing the performance of Rust with other languages focus on rarely used superficial algorithms, leading to somewhat inconclusive results. In this work, we conduct a comparative performance benchmark of Rust and C++ using commonly used algorithms and data structures rather than exotic ones. Our evaluation shows that the overall performance of Rust is similar to C++, with only minor disadvantage. We also demonstrate that in some Rust routines are slightly faster than the ones of C++.
\end{abstract}


\section{Introduction}

In this work, we measure the performance of C++ and Rust using selected commonly used algorithms, operations, and abstract data structures (ADS) implemented in the standard libraries of these languages. Merge Sort and Insertion Sort are among most popular and well-researched classic algorithms in Computer Science. However, some performance aspects of these algorithms still remain unclear. In this work, we fill this gap by comparing the performance of Insertion and Merge Sort under different sizes of input vectors using C++ and Rust implementations. The evaluation shows that: a) the size of input vector at which Merge Sort begins outperforming Insertion Sort depends upon implementation and optimization; b) Rust outperforms C++ in Merge Sort, while C++ outperforms Rust in Insertion Sort.

It has been known that Insertion Sort outperforms Merge Sort with small data series, which led to creation of Hybrid Sort that chooses which algorithm to apply depending on the size of the current partition. However, the specific optimal parameters of Hybrid sort are uncertain. In this work, we conduct an empirical study to shed light on the range of optimal partition size, denoted $K$, for the hybrid Merge+Insertion sort. For evaluation, we implement and run multiple tests using C++ and Rust programming languages. The evaluation shows that: a) the performance profile of Hybrid Sort implemented in C++ is different from the one implemented in Rust, including a significant difference in optimal range of $K$; b) Hybrid sort outperforms both Merge Sort and Insertion Sort --- significantly in C++ and slightly in Rust; c) C++ optimized implementation of Hybrid Sort outperforms the one in Rust.

Dictionary data structures allow for fast acquisition of mapped values by keys. However, the time complexity of modification of these data structures involves some uncertainty. Specifically, the insertions and deletions are believed to be faster in hash maps compared to balanced binary trees, but the exact difference is not well-established. In this work, we fill this gap by measuring the performance of dictionary-like data structures in three orthogonal dimensions: 1) abstract data structure (ADS) in question; 2) modifying operation applied; and 3) programming language used for implementation. The results show that: a) hash maps outperform binary trees in deletion and insertion; b) the performance of deletions is similar or faster than of insertions, depending on implementation; and c) C++ implementation performs these operations faster than Rust implementation.

\subsection{Insertion Sort and Merge Sort}

It is known that Insertion Sort is faster than Merge Sort on smaller data series~\cite{cse830,cormen2009introduction,skiena1998algorithm}. However, the graphs of their respective asymptotic time complexity functions, $y = x^2$ and $y = x \cdot log_2(x)$, do not exhibit any ``postponed crossing'', which suggests that the size of the input vector at which Merge Sort begins outperforming Insertion Sort must be very small.







\section{Methods}

\subsection{Merge Sort vs. Insertion Sort}

We implement Merge Sort and Insertion Sort using open-source algorithms provided by Dr. Charles Ofria at \url{https://github.com/mercere99/CSE-830}. For both Rust and C++ implementations, We use semantically identical algorithms. For the sorted data structure we use the standard library vectors of double precision floating point numbers: \texttt{vector<double>} in C++ and \texttt{Vec<f64>} in Rust. We used the same computer\footnote{AMD Ryzen Threadripper 2950x (16 cores), 72 Gb RAM, Ubuntu 20.04, with GNU g++ 9.3.0 as C++ compiler, and rustc 1.47.0 as Rust compiler.} under the same background load to perform the following 6 measurements:

\begin{itemize}
    \item Merge sort in optimized C++ build\footnote{GNU \texttt{g++} with \texttt{-O2} flag.};
    \item Merge sort in non-optimized C++ build\footnote{GNU \texttt{g++} without any optimization or debugging flags.};
    \item Insertion sort in optimized C++ build;
    \item Insertion sort in non-optimized C++ build;
    \item Merge sort in optimized Rust build\footnote{\texttt{cargo build} with \texttt{--release} flag.};
    \item Insertion sort in optimized Rust build.
\end{itemize}

For each measurement, we run 10,000 probes, and then calculate mean averages from these probes, which is used as the result of the measurement. For each measurement, we generate a vector with the size between 25 and 1,000 with the step of 25, i.e, $\{25, 50, 75, ..., 950, 975, 1000\}$. For each measurement, we create a new vector and populate it with random double-precision floating point numbers from the range $[1.0, 65536.0]$. For each vector size, we repeat the same procedure 10,000 times. The generation of random numbers and population of each vector are excluded from the delay measurement. In other words, for each run, only sorting delay is measured. The full implementation of the benchmark is available here: \url{https://github.com/nick-ivanov/rust-cpp-benchmark}.


\subsection{Hybrid Sort, Merge Sort, and Insertion Sort}

We implement Hybrid Sort, Merge Sort, and Insertion Sort using open-source algorithms provided by Dr. Charles Ofria at \url{https://github.com/mercere99/CSE-830}. For both Rust and C++ implementations we use semantically identical algorithms. For the sorted data structure we use the standard library vectors of double precision floating point numbers: \texttt{vector<double>} in C++ and \texttt{Vec<f64>} in Rust. We use the same computer\footnote{AMD Ryzen Threadripper 2950x (16 cores), 72 Gb RAM, Ubuntu 20.04, with GNU g++ 9.3.0 as C++ compiler, and rustc 1.47.0 as Rust compiler.} under the same background load to perform the following 16 measurements:

\begin{itemize}
    \item C++ Hybrid Sort with $K=16$\footnote{For all C++ builds we use GNU \texttt{g++} with \texttt{-O2} flag.};
    \item C++ Hybrid Sort with $K=32$;
    \item C++ Hybrid Sort with $K=64$;
    \item C++ Hybrid Sort with $K=128$;
    \item C++ Hybrid Sort with $K=256$;
    \item C++ Hybrid Sort with $K=512$;
    \item Rust Hybrid Sort with $K=16$\footnote{For all Rust builds we use \texttt{cargo build} with \texttt{--release} flag.};
    \item Rust Hybrid Sort with $K=32$;
    \item Rust Hybrid Sort with $K=64$;
    \item Rust Hybrid Sort with $K=128$;
    \item Rust Hybrid Sort with $K=256$;
    \item Rust Hybrid Sort with $K=512$;
    \item C++ Merge Sort;
    \item C++ Insertion Sort;
    \item Rust Merge Sort;
    \item Rust Insertion Sort.
\end{itemize}

Using these measurements, we perform the following 5 comparisons:

\begin{itemize}
    \item Hybrid Sort with 6 different values of $K$ measured in C++ implementation;
    \item Hybrid Sort with 6 different values of $K$ measured in Rust implementation;
    \item Hybrid Sort versus Merge Sort versus Insertion Sort in C++;
    \item Hybrid Sort versus Merge Sort versus Insertion Sort in Rust;
    \item Hybrid Sort in C++ versus Hybrid Sort in Rust.
\end{itemize}

For each measurement, we run 10,000 probes, and then calculate a mean average from these probes, which is used as the result of the measurement. For each probe, we generate a vector with the size between 250 and 10,000 with the step of 250, i.e, $\{250, 500, 705, ..., 9500, 9750, 10000\}$. For each measurement, we create a new vector and populate it with random double-precision floating point numbers from the range $[1.0, 65536.0]$. The generation of random numbers and population of each vector are excluded from delay measurement. In other words, for each run, only sorting delay is measured. The full implementation of the benchmark is available here: \url{https://github.com/nick-ivanov/rust-cpp-benchmark}.

\subsection{Dictionary and Binary Tree Insertion and Deletion}

We implement dictionary insertion for both hashmap and binary tree using loop-based addition of elements into dictionaries from two pre-generated vectors: one with unsigned 64-bit integer keys, and another with double-precision floating point random values. For dictionary deletion, we use a loop-based one-by-one deletion of all the elements using the vector of keys generated for insertion. For both Rust and C++ implementations, we use semantically identical algorithms for insertion and deletion. For the dictionary data structures we use the standard library implementations of \texttt{multimap<uint64\_t,double>} and \texttt{unordered\_multimap<uint64\_t,double>} in C++. For Rust dictionary data structures, we use \texttt{BTreeMap<u64,f64>} and \texttt{HashMap<u64,f64>}. Deletions and insertions are timed separately. We used the same computer\footnote{AMD Ryzen Threadripper 2950x (16 cores), 72 Gb RAM, Ubuntu 20.04, with GNU g++ 9.3.0 as C++ compiler, and rustc 1.47.0 as Rust compiler.} under the same background load to perform the following 8 measurements:

\begin{itemize}
    \item C++ hashmap insertion\footnote{For all C++ builds we use GNU \texttt{g++} with \texttt{-O2} flag.};
    \item C++ hashmap deletion;
    \item Rust hashmap insertion\footnote{For all Rust builds we use \texttt{cargo build} with \texttt{--release} flag.};
    \item Rust hashmap deletion;
    \item C++ binary tree insertion;
    \item C++ binary tree deletion;
    \item Rust binary tree insertion;
    \item Rust binary tree deletion.
\end{itemize}

We compare the performance of the operations with the data structures within the following three orthogonal dimensions:

\begin{itemize}
    \item \textbf{ADS:} \textit{what is faster, hashmap or balanced binary tree?}
    \item \textbf{operation:} \textit{what is the performance of the insertion compared to deletion?}
    \item \textbf{language:} \textit{which standard library implementation performs the above operations faster: in C++ or Rust?}
\end{itemize}

Below are 8 comparisons over the three dimensions that we make:

\begin{itemize}
    \item hashmap insertion versus balanced binary tree insertion in optimized C++ code;
    \item hashmap deletion versus balanced binary tree deletion in optimized C++ code;
    \item hashmap insertion versus balanced binary tree insertion in optimized Rust code;
    \item hashmap deletion versus balanced binary tree deletion in optimized Rust code;
    \item hashmap insertion in optimized C++ code versus hashmap insertion in optimized Rust code;
    \item hashmap deletion in optimized C++ code versus hashmap deletion in optimized Rust code;
    \item balanced binary tree insertion in optimized C++ code versus balanced binary tree insertion in optimized Rust code;
    \item balanced binary tree deletion in optimized C++ code versus balanced binary tree deletion in optimized Rust code.
\end{itemize}

We perform measurements of 100 different dictionary sizes from 100 to 10,000, with a step of 100, i,e., $100, 200, ..., 9,900, 10,000$. For each measurement, we run 10,000 probes, and then calculate a mean average from each probe, which is used as the result of the measurement. For each probe, we create two vectors of the same size:

\begin{itemize}
    \item vector of unsigned 64-bit integer random keys from the range of $[0, 2^{64}-1]$;
    \item vector of double precision random floating point values from the range of $[1.0, 65536.0]$.
\end{itemize}

Then, the generated keys and values are inserted in the dictionary. Then the same keys are used to remove these elements from the dictionary one by one. The time of both operations is measured individually. The generation of random numbers and population of each vector are excluded from delay measurement. In other words, for each run, only insertion or deletion delay is measured. The full implementation of the benchmark is available here: \url{https://github.com/nick-ivanov/rust-cpp-benchmark}

\section{Results}

\subsection{Sorting Algorithm Comparison in C++}
\begin{figure}
    \centering
    \includegraphics[width=\linewidth]{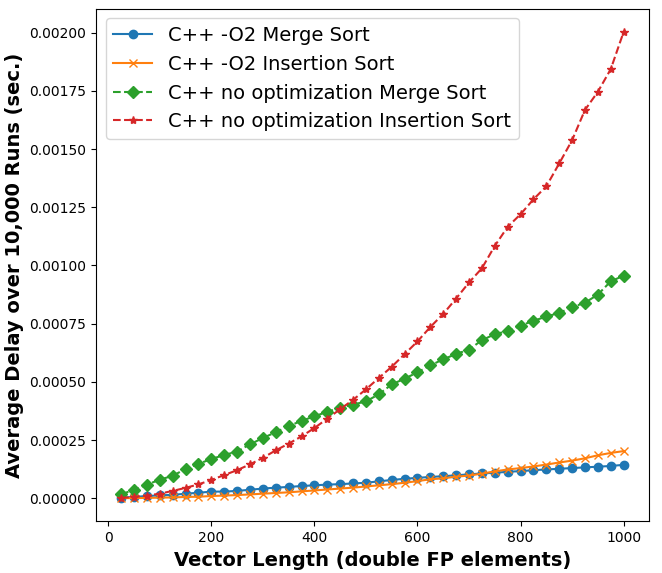}
    \caption{Comparison of performance between Insertion Sort and Merge Sort in optimized and non-optimized C++ implementations.}
    \label{fig:plot11}
\end{figure}

Figure~\ref{fig:plot11} shows that the size of the vector at which Merge Sort begins outperforming Insertion Sort depends upon implementation. The experiment shows that in non-optimized C++ build, the pivot point is between 400 and 500 elements, whereas if the same code is compiled with \texttt{-O2} parameter, the pivot moves to the range between 700 and 800 elements.

\subsection{Sorting Algorithm Comparison in Rust}
\begin{figure}
    \centering
    \includegraphics[width=\linewidth]{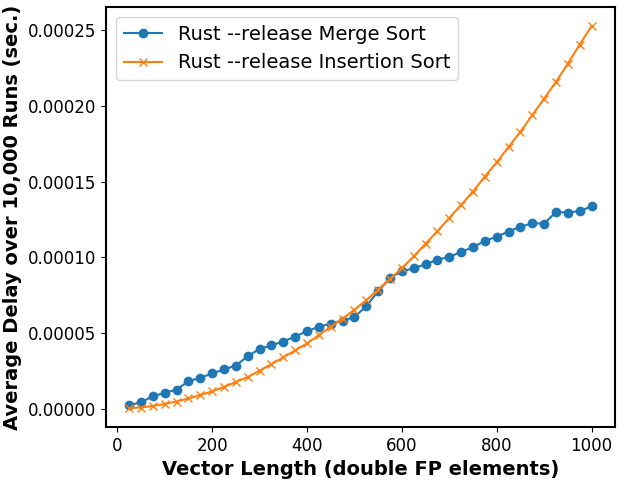}
    \caption{Comparison of performance between Insertion Sort and Merge Sort in optimized Rust implementation.}
    \label{fig:plot12}
\end{figure}

As it is further confirmed by Figure~\ref{fig:plot12}, the length of the vector at which Merge Sort becomes faster than Insertion Sort is indeed implementation-dependent. The Rust implementation of the same algorithm demonstrates two crossings of the performance graphs --- one between 400 and 500, and another one between 500 and 600 elements.

\subsection{Merge Sort in C++ and Rust}
\begin{figure}
    \centering
    \includegraphics[width=\linewidth]{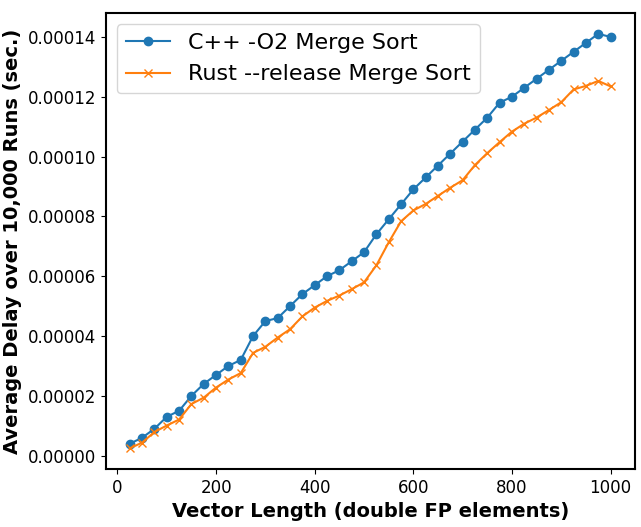}
    \caption{Comparison of performance between optimized implementation of Merge Sort in C++ and optimized implementation of Merge Sort in Rust.}
    \label{fig:plot13}
\end{figure}

Figure~\ref{fig:plot13} shows that Rust outperforms C++ in Merge Sort for each vector length within the tested range.

\subsection{Insertion Sort in C++ and Rust}
\begin{figure}
    \centering
    \includegraphics[width=\linewidth]{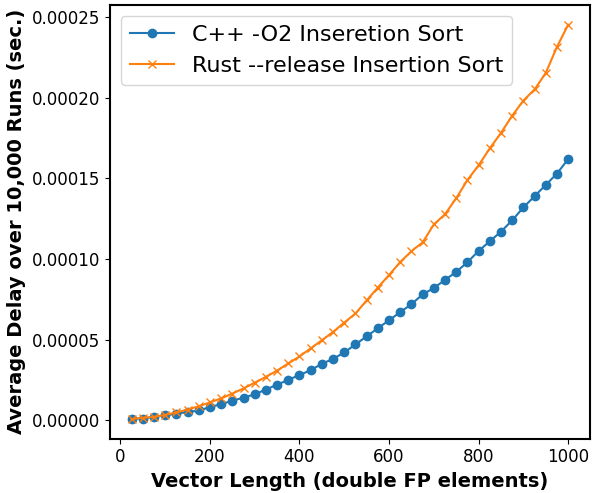}
    \caption{Comparison of performance between optimized implementation of Insertion Sort in C++ and optimized implementation of Insertion Sort in Rust.}
    \label{fig:plot14}
\end{figure}

Figure~\ref{fig:plot14} demonstrates that C++ outperforms Rust in Insertion sort for each vector length in the tested range.

\subsection{Hybrid Sort}

\subsubsection{Hybrid Sort in C++}
\begin{figure}
    \centering
    \includegraphics[width=\linewidth]{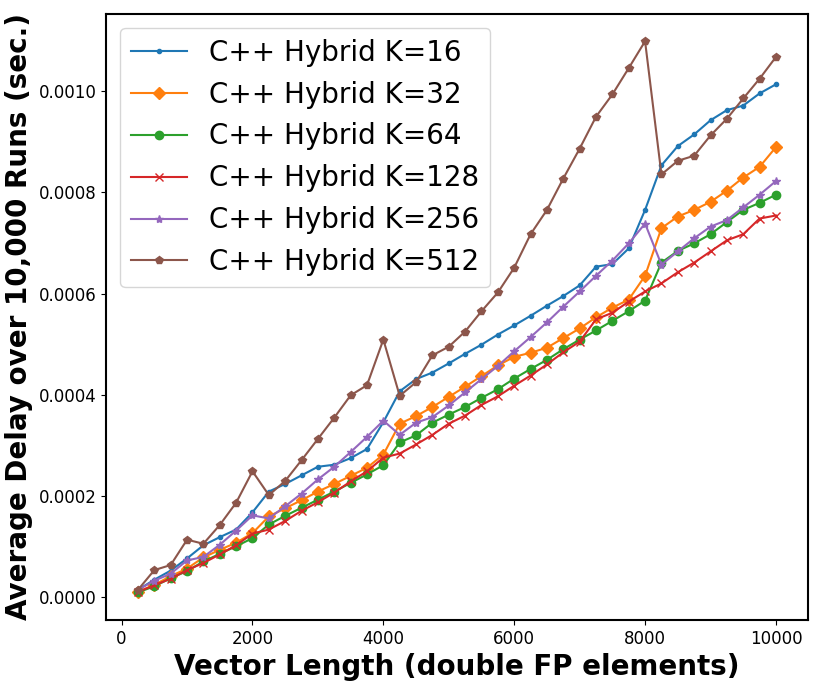}
    \caption{Comparison of performance between hybrid sort algorithm with different thresholds in optimized C++ implementation.}
    \label{fig:plot21}
\end{figure}

Figure~\ref{fig:plot21} shows that Hybrid Sort in C++ varies depending on the value of $K$. However, the implementations with different values of $K$ perform differently within different ranges. Although it is not easy to establish the top performer, the value $K=128$ exhibits most minimums and the absence of significant spikes, and therefore chosen to be the best parameter among the 5 others.

\subsubsection{Hybrid Sort in Rust}
\begin{figure}
    \centering
    \includegraphics[width=\linewidth]{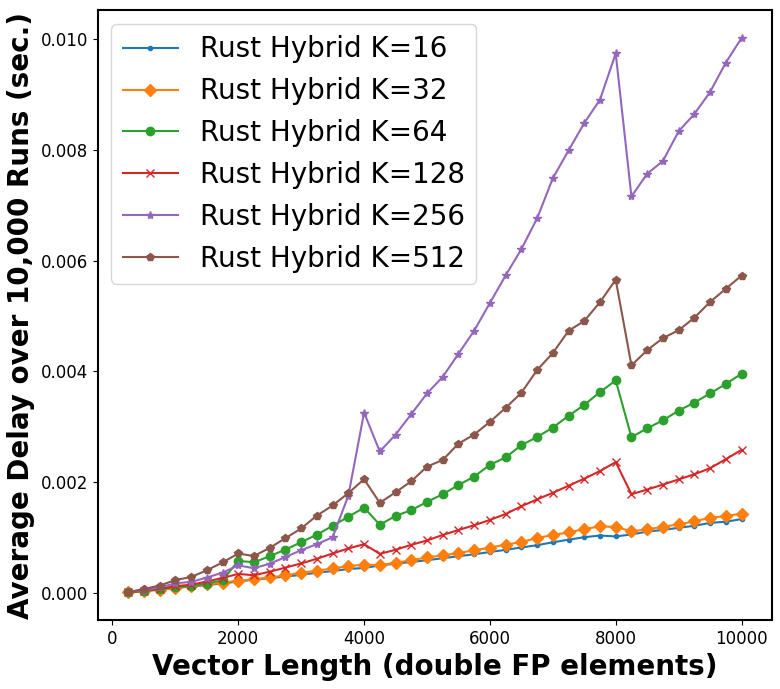}
    \caption{Comparison of performance between hybrid sort algorithm with different thresholds in optimized Rust implementation.}
    \label{fig:plot22}
\end{figure}

As per Figure~\ref{fig:plot22}, the performance of Hybrid Sort in Rust demonstrates stronger dependence upon the value of $K$, with the exception of $K=512$, which shows a significant reduction in performance starting in the range between 3,000 and 4,000 items. The overall leader is easy to identify for Rust ($K=16$). Unlike C++, in which very low and very high values of $K$ are among the worst, in Rust the smallest value of $K$ is the best, which further testifies that the performance of Hybrid sort varies depending on implementation.

\subsubsection{Hybrid Sort Compared to Other Sorts in C++}
\begin{figure}
    \centering
    \includegraphics[width=\linewidth]{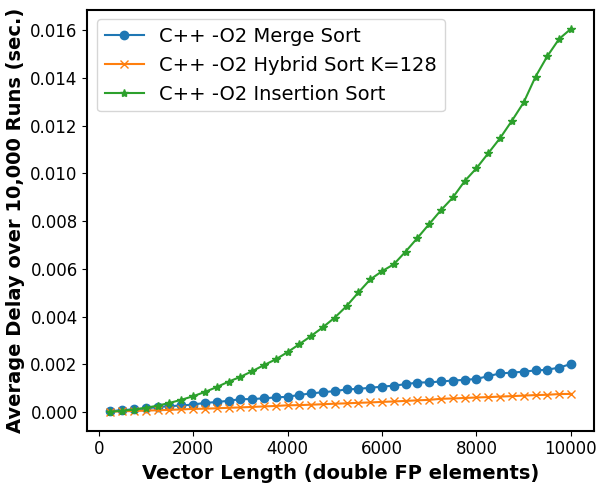}
    \caption{Comparison of performance between Merge Sort, Insertion Sort, and Hybrid Sort with the threshold K=128 in optimized C++ code.}
    \label{fig:plot23}
\end{figure}

As shown in Figure~\ref{fig:plot23}, Hybrid Sort outperforms Merge Sort and Insertion Sort in C++, which is expected. The parameter $K$ chosen for this comparison is the previously determined best performer for C++, i.e, $K=128$.

\subsubsection{Hybrid Sort Compared to Other Sorts in Rust}
\begin{figure}
    \centering
    \includegraphics[width=\linewidth]{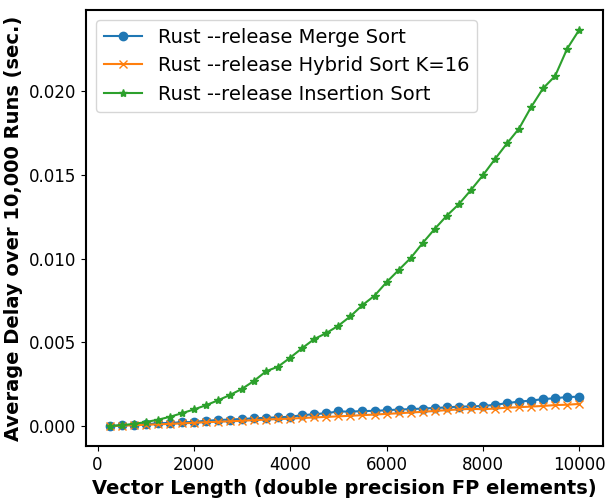}
    \caption{Comparison of performance between Merge Sort, Insertion Sort, and Hybrid Sort with the threshold K=16 in optimized Rust code.}
    \label{fig:plot24}
\end{figure}

Figure~\ref{fig:plot24} shows that Hybrid Sort outperforms Merge Sort and Insertion Sort in Rust, which is expected. However, the performance difference between Merge Sort and Hybrid Sort in Rust is very small. The parameter $K$ chosen for this comparison is the previously determined best performer for Rust, i.e, $K=16$.

\subsubsection{Hybrid Sort in C++ and Rust}
\begin{figure}
    \centering
    \includegraphics[width=\linewidth]{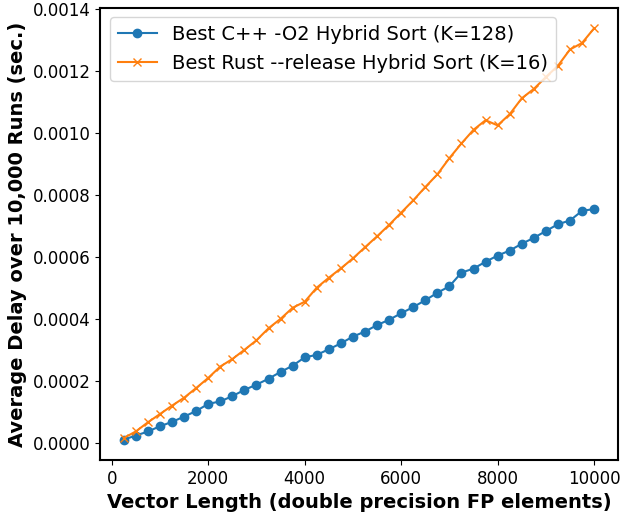}
    \caption{Comparison of performance between best performing Hybrid Sort in optimized C++ implementation with the best performing Hybrid Sort in optimized Rust implementation.}
    \label{fig:plot25}
\end{figure}

Figure~\ref{fig:plot25} compares performance of Hybrid Sort in C++ and Rust with their respective best performing values, i.e., $K_{C++}=128$, and $K_{Rust}=16$. The result shows superior performance of C++ for each size of the input vector in the tested range.

\subsection{Dictionary Operations}

\subsubsection{C++ Insertions}
\begin{figure}
    \centering
    \includegraphics[width=\linewidth]{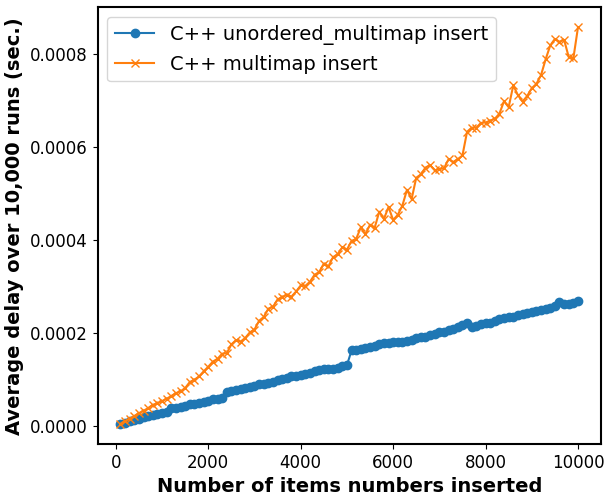}
    \caption{Comparison of performance of hashmap insertion with balanced binary tree insertion in optimized C++ code.}
    \label{fig:plot31}
\end{figure}

Figure~\ref{fig:plot31} shows that in optimized C++ code the insertion operation in a hashmap is always faster than in a balanced binary tree, which is expected~\cite{cse830,skiena1998algorithm,cormen2009introduction}. The insertion in both the data structures is demonstrating a visibly linear pattern with relatively small fluctuations.

\subsubsection{C++ Deletions}
\begin{figure}
    \centering
    \includegraphics[width=\linewidth]{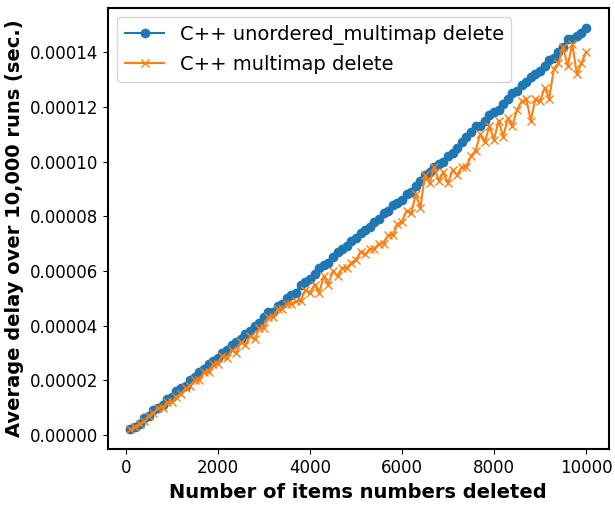}
    \caption{Comparison of performance of hashmap deletion with balanced binary tree deletion in optimized C++ code.}
    \label{fig:plot32}
\end{figure}

As we can see in Figure~\ref{fig:plot32}, the performance of hashmap deletions in C++ is only slightly higher than of balanced binary tree. However, it is evident that overall dictionary deletions in C++ are faster than insertions.

\subsubsection{Rust Insertions}
\begin{figure}
    \centering
    \includegraphics[width=\linewidth]{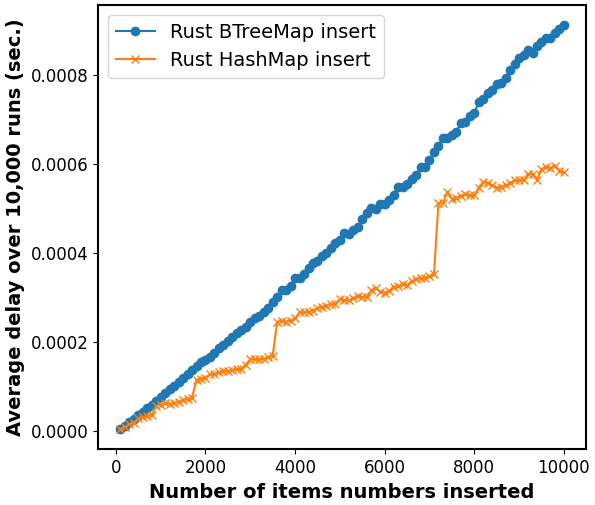}
    \caption{Comparison of performance of hashmap insertion with balanced binary tree insertion in optimized Rust code.}
    \label{fig:plot33}
\end{figure}

Figure~\ref{fig:plot33} shows that in Rust, like in C++, the insertion operation is faster for hashmap. However, the performance difference betwen hashmap and binary tree in Rust is smaller than in C++.

\subsubsection{Rust Deletions}
\begin{figure}
    \centering
    \includegraphics[width=\linewidth]{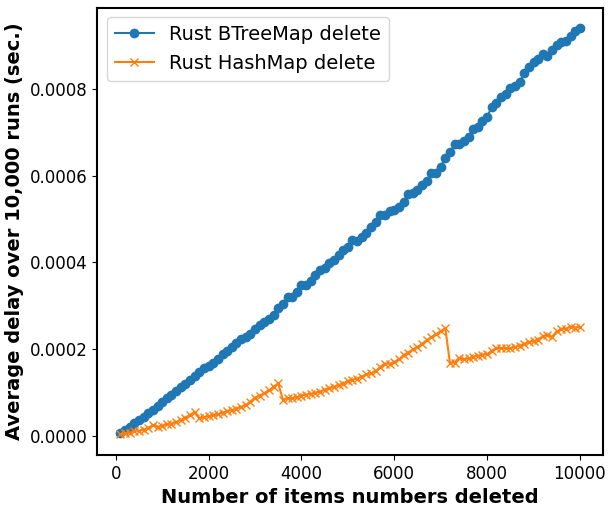}
    \caption{Comparison of performance of hashmap deletion with balanced binary tree deletion in optimized Rust code.}
    \label{fig:plot34}
\end{figure}

As shown in Figure~\ref{fig:plot34}, the performance difference between deletion operations in hashmap and binary tree in Rust is in favor of hashmap and is significantly larger than in C++.

\subsubsection{Hashmap Insertions by Language}
\begin{figure}
    \centering
    \includegraphics[width=\linewidth]{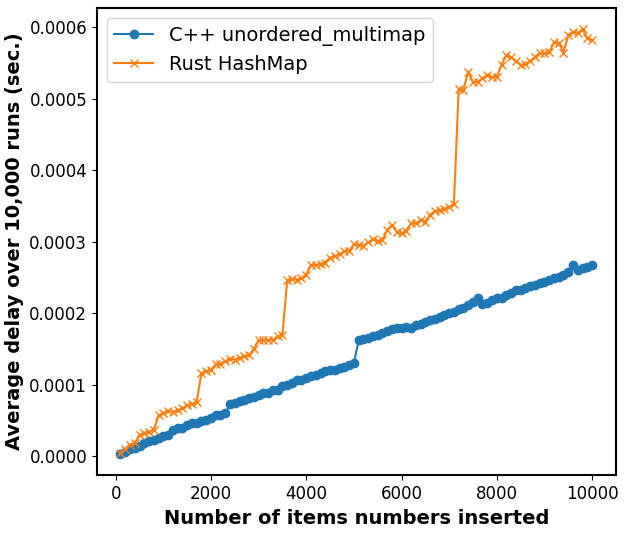}
    \caption{Comparison of performance of hashmap insertion in optimized C++ code with hashmap insertion in optimized Rust code.}
    \label{fig:plot35}
\end{figure}

Figure~\ref{fig:plot35} shows that hashmap insertion operations in C++ are faster than in Rust for the entire test set.

\subsubsection{Hashmap Deletions by Language}
\begin{figure}
    \centering
    \includegraphics[width=\linewidth]{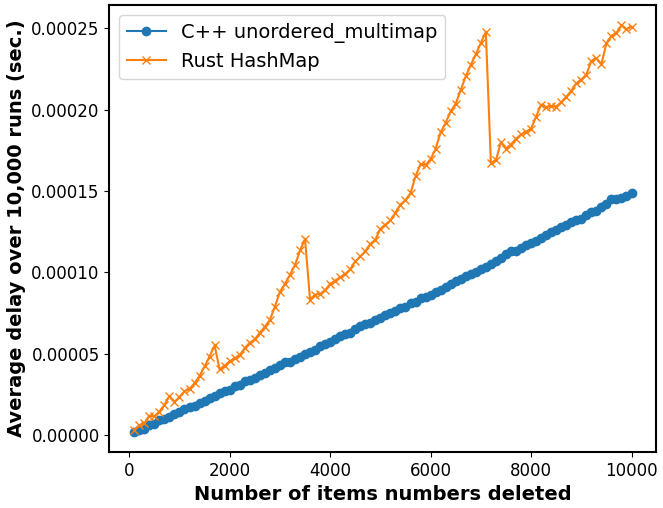}
    \caption{Comparison of performance of hashmap deletion in optimized C++ code with hashmap deletion in optimized Rust code.}
    \label{fig:plot36}
\end{figure}

Figure~\ref{fig:plot36} shows that hashmap deletion operations in C++ are faster than in Rust for the entire test set.

\subsubsection{Binary Tree Insertions by Language}
\begin{figure}
    \centering
    \includegraphics[width=\linewidth]{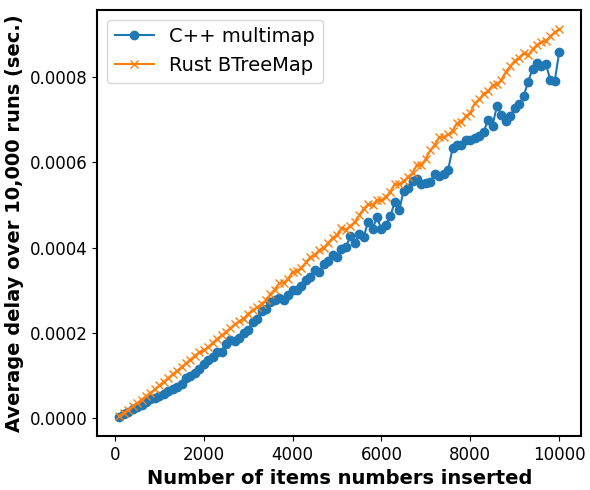}
    \caption{Comparison of performance of balanced binary tree insertion in optimized C++ code with balanced binary tree insertion in optimized Rust code.}
    \label{fig:plot37}
\end{figure}

Figure~\ref{fig:plot37} shows that binary tree insertion operations in C++ are slightly faster than in Rust for the entire test set.

\subsubsection{Binary Tree Deletions by Language}
\begin{figure}
    \centering
    \includegraphics[width=\linewidth]{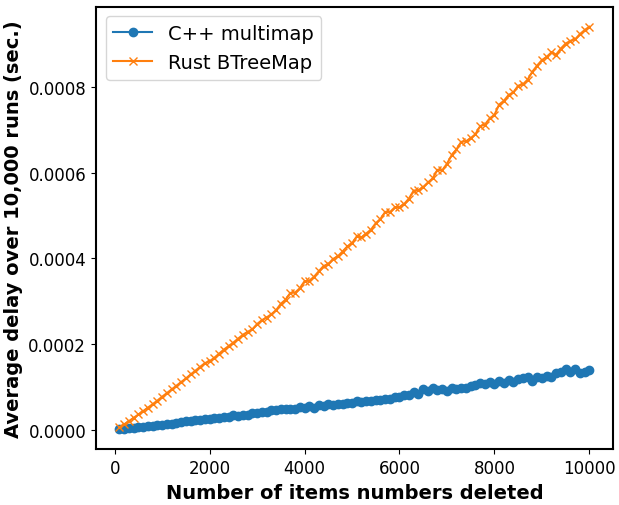}
    \caption{Comparison of performance of balanced binary tree deletion in optimized C++ code with balanced binary tree deletion in optimized Rust code.}
    \label{fig:plot38}
\end{figure}

Figure~\ref{fig:plot38} shows that binary tree deletion operations in C++ are faster than in Rust for the entire test set.

\section{Discussion}
The major limitation of this work is that the data for each probe is generated randomly. More research is needed to assess the performance of data sets with various non-random distributions common for the real-world data, such as partially sorted series and vectors with multiple repeated elements.

\section{Conclusions}
In this work, we demonstrated that: a) the size of input vector at which Merge Sort starts performing better than Insertion Sort depends on the implementation and optimization and is largely unpredictable; b) Rust outperforms C++ in Merge Sort, while C++ outperforms Rust in Insertion Sort. We further discovered that: a) the performance of Hybrid Sort implemented in C++ is different from the one implemented in Rust, with a significant difference in optimal range of $K$; b) Hybrid sort outperforms both Merge Sort and Insertion Sort, especially in C++; c) C++ optimized implementation of Hybrid Sort outperforms the one in Rust. Finally, we discovered that hashmaps indeed outperform binary trees in deletion and insertion. Also, we showed that and C++ implementation performs all insertion and deletion dictionary operations faster than Rust. Also, the study showed that the performance of deletions is similar to insertions in Rust, but faster than insertions in C++.



\end{document}